\documentstyle[epsfig,prl,preprint,aps]{revtex}

\begin{document}
\title{Experimental evidence of a phase transition to fully developed 
turbulence in a wake flow} 
\author{St.  L\"uck, J. Peinke} \address{Fachbereich Physik, Universit\"at Oldenburg} 
\author{R. Friedrich} \address{Institut f\"ur theoretische Physik Universit\"at Stuttgart} 
\date{\today} 
\maketitle
\begin{abstract}
The transition to fully developed turbulence of a wake behind a circular cylinder is 
investigated with respect to its statistics. In particular, we evaluated the 
probability density functions of velocity increments on 
different length scales $r$. Evidence is presented that the $r$-dependence of 
the velocity increments can be taken as Markov processes in the far 
field, as well as, in the near field of the cylinder wake.  With the 
estimation of the deterministic part of these Markov processes, as a 
function of the distance from the cylinder, we are able to set the 
transition to fully developed turbulence in analogy with a phase 
transition.  We propose that the appearing order parameter corresponds 
to the presence of large scale coherent structures close to the 
cylinder.
\end{abstract}

\section{Introduction}

In recent years considerable progress has been performed to understand 
the statistical features of fully developed, local isotropic 
turbulence \cite{turb}.  Special interest has been addressed to 
understand intermittency effects of small scale velocity fluctuations 
characterized by the velocity increments $u_{r}(x) := u(x+r)-u(x)$ at 
a scale $r$.  For most real flows these results are only 
applicable for small well defined regions of the flow, which may 
be regarded as local isotropic.  A remaining challenge is to find out 
how theses concepts can help to understand real flows  
which are not fully developed or not homogeneous and isotropic \cite{realturb}.

The common method to characterize the disorder of fully developed 
local isotropic turbulence is to investigate the scale evolution of 
the probability density functions (pdf), $P_{r}(u_{r})$, either 
directly or by means of their moments $<u_{r}^n> = \int u_{r}^n \, 
P(u_{r}) \,du_{r}$.  Recently, it was found that 
this $r$ evolution can be related to a Markov process \cite{Markov}.  
The Markovian properties can be evaluated thoroughly by investigating 
the joint and conditional pdfs, $P(u_{r2},r2;u_{r1},r1)$ and 
$P(u_{r2},r2|u_{r1},r1)$, respectively \cite{zeller}.  From the 
conditional pdfs one can extract the stochastic equations, namely, the 
Kramers-Moyal expansion for the $r$ evolution of $P_{r}$ and the 
Langevin equation for $u_{r}$ \cite{Risken}.  This method provides a 
statistically more complete description of turbulence and furthermore 
assumptions, like scaling, are not needed, but can be evaluated accurately 
\cite{Markov,scaling}.

In this work we present measurements of a turbulent flow behind a 
circular cylinder.  The stochastic content of the velocity field as a 
function of the distance to the cylinder is investigated using the 
above mentioned Markovian approach.  The main result, presented in this 
paper, is the finding of a phase transition like behavior to the state 
of fully developed turbulence.  This 
phase transition characterizes the disappearance of the Karman vortices 
with respect to two parameters: the distance to the cylinder and the scale 
$r$.

In the following we describe first the experimental set up.  The 
measurements of longitudinal and transversal velocities are analyzed 
with respect to the $r$ dependent pdfs.  Subsequently a test of Markov 
properties is presented.  From the conditional pdf the first moment 
$M^{(1)}$ is evaluated.  The $M^{(1)}$ coefficient reflects the 
deterministic part in the $r$-evolution of the Markov process and can be 
taken to define an order parameter.

\section{experiment}

Our work is based on hot-wire velocity measurements performed in a 
wake flow generated behind a circular cylinder inserted in a wind 
tunnel.  Cylinders with two diameters $d$ of 2 cm and 5 cm were used.  
The wind tunnel \cite{windtunnel} used has the following parameters: 
cross section 1.6m x 1.8m; length of the measuring section 2m; 
velocity 25m/s; residual turbulence level below 0.1 \%. To measure 
longitudinal and transversal components of the local velocity we used 
x-wire probes (Dantec 55P71), placed at several distances, $D$, 
between 8 and 100 diameters of the cylinder. The spatial resolution 
of the probes is about 1.5 mm.

From the measurements the following characteristic lengths were 
evaluated: the integral length, defined by the autocorrelation 
function, which varied between 10 cm and 30 cm depending on the 
cylinder used and location of the probe; the Kolmogorov length, was 
about 0.1mm; the Taylor length scale about 2.0 mm.  Thus we see that 
our measurement resolved at least the turbulent structures down to the 
Taylor length scales.  (Note, these lengths could be calculated 
precisely only for distances above 40 cylinder diameters.)  The 
Reynolds numbers of these two flow situations were $R_{\lambda} =$ 250 
and 650.  Each time series consists of $10^7$ data points, and was 
sampled with a frequency corresponding to about one Kolmogorov length.  
To obtain the spatial variation the Taylor hypothesis of frozen 
turbulence was used.

\section{results}

To investigate the disorder of the turbulent field the velocity 
increments for different scales $r$ and at different measuring points 
$D$ were calculated.  Exemplary sequences of resulting pdfs are shown 
in Fig.1 for the transversal velocity component.  In Fig.  1a the well 
known intermittency effect of isotropic turbulence is seen.  At large 
scales nearly Gaussian distributions are present which become more and 
more intermittent (having heavy tailed wings) as the scale $r$ 
approaches the Kolmogorov length.  Coming closer to the cylinder a 
structural change is found.  Most remarkably a double hump pdf emerges 
for large $r$.  This structure reflects the fact that two finite 
values of the velocity increment are most probable.  We interpret this 
as the result of counterrotating vortices passing over the detector.  
It should be noted that this effect was always found for the 
transversal velocity components.  This is in consistency with the 
geometric features of vortices elongated parallel to the cylinder axis 
(Karman vortices).  For small scales the humps vanish and the pdfs 
become similar to the isotropic ones.

\section{Markov Process}

Based on the findings that the evolution of the pdfs with $r$ for the 
case of fully developed turbulence can be described by a Fokker-Planck 
equation \cite{Markov,Renner}, we apply the Markov analysis to the non 
fully developed states close to the cylinder.  The basic quantity to 
be evaluated is the conditional pdf $P(u_{r2},r2|u_{r1},r1)$, where, 
$r2<r1$, and $u_{r2}$ is nested into $u_{r1}$ by a midpoint 
construction.  To verify the Markovian property, we evaluate the 
Chapman-Kolmogorov equation, c.f.  \cite{Risken}
\begin{equation}
	P(u_{r2},r2|u_{r1},r1) = 
	\int_{-\infty}^{\infty} \, P(u_{r2},r2|u_{rx},rx) 
	P(u_{rx},rx|u_{r1},r1) \, d u_{rx} ,
	\label{ChapKol}
\end{equation}
 where $r2 < rx < r1$.  The validity of this equation was examined for 
 many different pairs of $(r1,r2)$.  As a new result, we found that 
 equation (\ref{ChapKol}) also holds in the vicinity of the cylinder, 
 i.e.  in the non developed case of turbulence.  For illustration see 
 Figure 2; in part a the integrated conditional pdf (rhs of 
 (\ref{ChapKol})) and the directly evaluated pdf (lhs of 
 (\ref{ChapKol})) are shown by superimposed contour plots.  In 
 figurepart b three exemplary cut through these three dimensional 
 presentations are shown.  The quality of the validity of 
 (\ref{ChapKol}) can be seen from the proximity of the contour lines, 
 or by the agreements of the conditional pdfs, represented by open and 
 bold symbols \cite{comment}.  Based on this result we treat the 
 evolution of the statistics with the scale $r$ as a Markov process in 
 $r$.  Thus the evolution of the pdf $P_{r}(u_{r})$ is described by 
 the partial differential equation called Kramers-Moyal expansion 
 \cite{Risken}:
\begin{equation}
   -\frac{d}{dr}P(u_{r},r)=\sum_{k=1}^{\infty} [-\frac{\partial}
       {\partial u_{r}}]^k D^{(k)}(u_{r},r)P(u_{r},r)
   \label{KraMol}
\end{equation}
with the coefficients 
\begin{eqnarray}
	M^{(k)}(u_{r},r,\delta) & := & \frac{1}{k!} \frac{1}{\delta} 
                    \int du_{rx} \; (u_{rx}-u_{r})^k \; p(u_{rx}, rx | 
                    u_{r}, r)\\
	D^{(k)}(u_{r}, r) &:=& \lim_{\delta\rightarrow0} M^{(k)}(u_{r},r,\delta),
   \label{KraMolCoeff}
\end{eqnarray}
where $\delta=r-rx$.  Notice, having once evaluated the conditional 
pdfs, these so called Kramers Moyal (KM) coefficients can be estimated 
directly from the data without any additional assumption.  For our 
purpose it is sufficient to consider the $M^{(k)}$ for a small length 
of $\delta\approx 2\eta$.  The physical interpretation of the KM 
coefficients is the following: $D^{(1)}$ describes the deterministic 
evolution and is called drift term.  $D^{(k)}$, for $k \geq 2$ reflect 
the influence of the noise.  $D^{(2)}$ is called diffusion term.  In 
the case of non Gaussian noise the higher order KM coefficients 
($k>2$) become non zero.

We found that the structural change of the pdfs described above (see 
fig.  1) is mainly given by $M^{(1)}$.  As shown in figure 3, we find 
that close to the cylinder the form of the $M^{(1)}$ changes from a 
linear $u_{r}$-dependence at small scales to a 3rd order polynomial 
behavior.  From the corresponding Langevin equation \cite{Risken} we 
know that the zeros of the drift term correspond to the fixed points 
of the deterministic dynamics.  Fixed points with negative slope 
belong to accumulation points, having the tendency to build up local 
humps in the pdf.  The change of the local slope of a fixed point (fig 
3) can be set into correspondence to a phase transition, c.f.  
\cite{phasetr,Risken}.  Note that in contrast to other models, where the 
process evolves in time, here, we stress on the evolution in the scale 
variable $r$.

The main point of our analysis is that we can determine the evolution 
equation in form of the KM coefficients.  This tool is much more 
sensitive than merely looking at the pdfs or its moments, because the 
pdfs reflect only the transient behavior due to the underlying 
evolution equation.  Thus it becomes clear that we are able to 
elaborate the phase transition even in the case where the double 
hump structure in the pdf may not be clearly visible.  We want to 
mention that these double hump behavior of the pdfs can well be 
reproduced by calculating the stationary solution of the corresponding 
Fokker Planck equation, using our measured KM coefficient $M^{(1)}$
\cite{Ich}.

Beside the spatial scale parameter $r$, the second parameter of the 
wake experiment is the distance of the probe to the cylinder.  As it 
is well known, with increasing the distance a transition to fully 
developed turbulence takes place, i.  e., the double hump structure 
vanishes.  To characterize the phase transition in this two 
dimensional parameter space more completely we performed the above 
mentioned data analysis at several distances.  As a criteria of a 
phase transition the local slope at $M^{(1)}(u_{r}, r, z)=0$ was 
determined.  The magnitude of this local slope is shown in figure 4 as 
a contour plot.  The dark colored region reflects the parameter space, 
where 3 zeros for $M^{(1)}$ are present, or where the local slope at 
$u_{r} = 0$ is positive.  This is the region where the new order 
parameter exists.  The critical line of the phase transition is marked 
by the bold black line.

\section{Discussion and Conclusion}

We have presented a new approach to characterize also the disorder of 
not fully developed turbulence.  The central aspect is that the 
disorder, described by velocity increments on different length 
scales, $r$, are set into the context of Markov processes evolving in $r$.  
Thus we can see how a given increment changes with decreasing $r$ due 
to deterministic and random forces.  Both forces can be estimated 
directly from the data sets via Kramers-Moyal coefficients of the 
conditional probabilities.  Most interestingly, we find significant 
changes in the deterministic force, the drift term, as one passes from 
non fully developed turbulence (close to the cylinder) into fully 
developed turbulence (far behind the cylinder).  In the far field the 
drift term causes a stable fixed point at $u_{r}= 0$, i.e.  the 
deterministic force causes a decrease of the magnitude of velocity 
increments as $r$ decreases.  Approaching the near field at large $r$ 
this fixed point becomes instable, i.e.  the slope of the drift term 
changes its sign at $u_{r}=0$.  In our one-dimensional analysis we 
find the appearance of two new stable (attracting) fixed point which 
are related to the double hump structure of the corresponding pdfs.  
This phenomenon may be set into relation with a phase transition, 
where the phase of the near field correspond to the existence of 
vortices.  As the distance to the cylinder is increased these large 
scale structures vanish.

Finally some critical remarks are presented, to show in which 
direction work should be done in future.  Visualizations indicate that 
even in the case of strong turbulence, the near field still resembles 
time periodic structures of counterrotationg vortex-like structures 
detaching from the cylinder, c.f.  \cite{Dyke}.  Theses time periodic 
large scale structures ask for a two-dimensional (two variable) 
modeling, in the sense of a noisy limit cycle.  This apparent 
contradiction to our one-variable analysis, has to be seen on the 
background of the signal treatment.  Applying to a time series the 
construction of increments (which represents a kind of high pass 
filter) the locality in time is lost.  Thus also coherences in time 
may get lost, at least as long as one investigates small scale 
statistics.  In this sense only a stochastic aspect of the 
counterrotating vortices is grasped.  The challenge of a more complete 
characterization of the near field structures will require, in our 
opinion, a combination of increment analysis and real time modeling of 
the velocity data.  At least for the ladder point a higher dimensional 
ansatz is required.  Nevertheless we have presented in this work clear 
evidence how methods and results obtained from the idealistic case of 
fully developed turbulence can be used to characterize also the 
statistics in the transition region of a wake flow.

Acknowledgment:
This work was supported by the DFG grant PE478/4. Furthermore we want 
to acknowledge the cooperation with the LSTM, namely, with T. Schenck, 
J. Jovanovic, F. Durst, as well as fruitful discussions with F. Chilla, Ch. 
Renner, B. Reisner and A. Tilgner.

%
%
\begin{figure}[ht]
  \begin{center}
    \epsfig{file=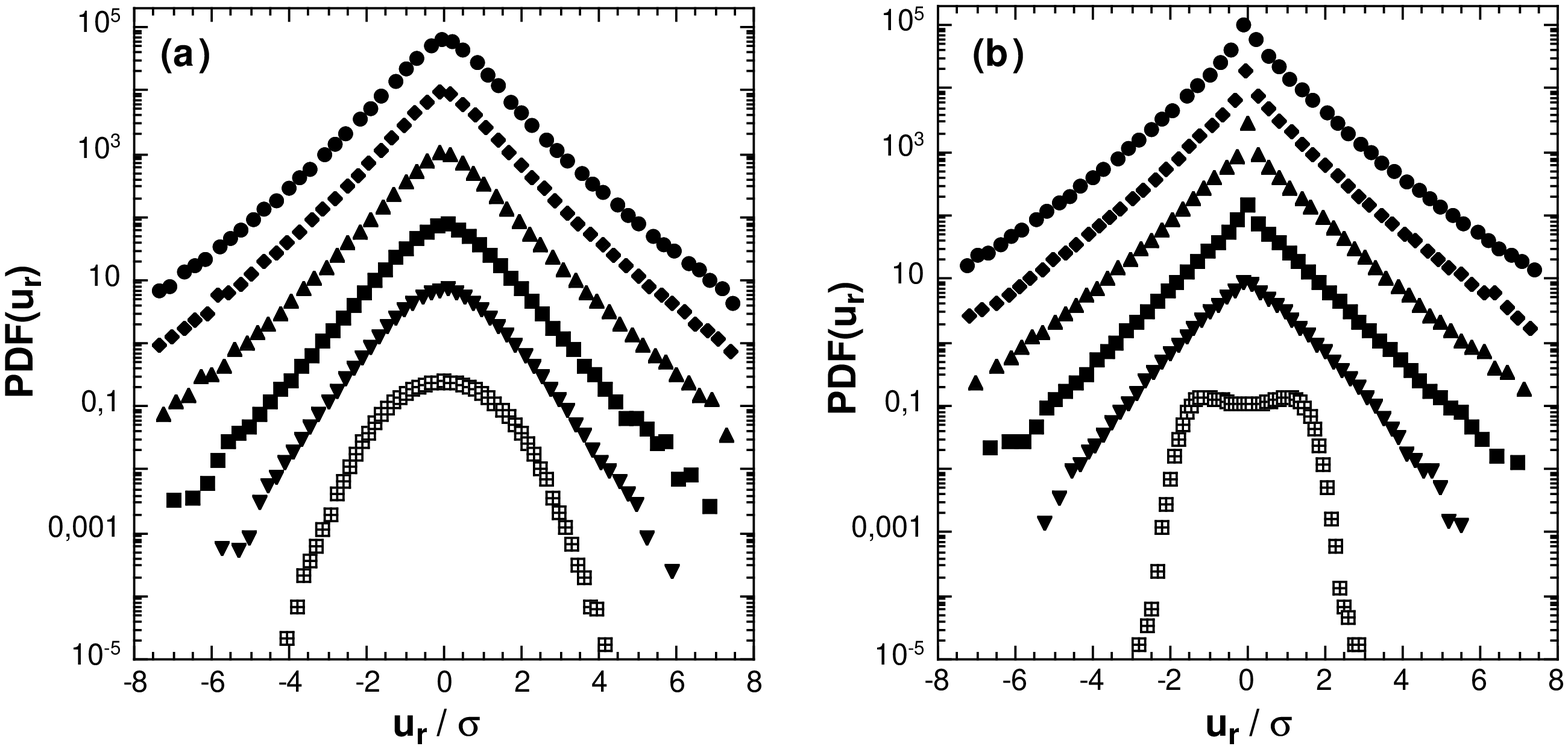, width=8.0cm}
  \end{center}
 \caption{Probability density functions for $r=0.1$ mm to $r=120$ mm (from 
 top to bottom) obtained from two data sets of transversal velocities, cylinder 
 diameter $d=5$ cm. a) fully developed turbulence (40 d); b) transition 
 region close to the cylinder (8 d). pdfs are shifted along the y 
 direction for clearness of the presentation.}
 \label{pdfs}
\end{figure}
%
%
\begin{figure}[ht]
    \begin{center}
       \epsfig{file=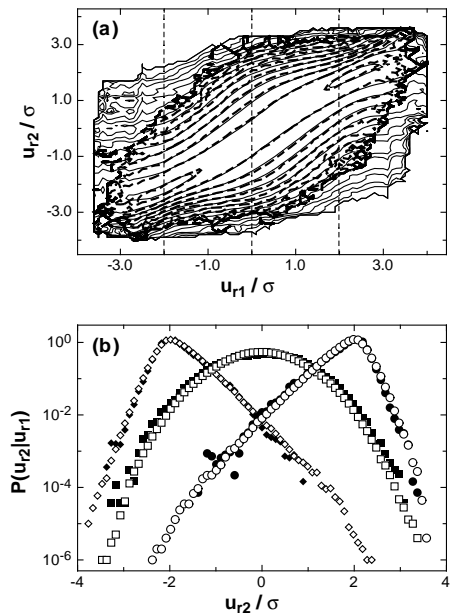,  width=6.0cm }
     \end{center}
 \caption{Verification of the Chapman Kolmogorov equation in the 
 transition region (D = 8d) for the cylinder with $d=5$ cm. 
 a) contour plot of the directly evaluated conditioned probability 
 distribution, presented as dashed lines, and numerically 
 integrated cond. pdf. (rhs of equation \ref{ChapKol}) represented by solid 
 lines (r1=10.5 cm, r2=12.9 cm). b) Corresponding cuts for selected $u_{r1}$ values (see marked 
 lines in a)). Bold symbols stand for the directly evaluated 
 conditional pdf and open symbols for the integrated conditional pdfs.}
 \label{ChapKolFig}
\end{figure}
%
%
\begin{figure}[ht]
  \begin{center}
    \epsfig{file=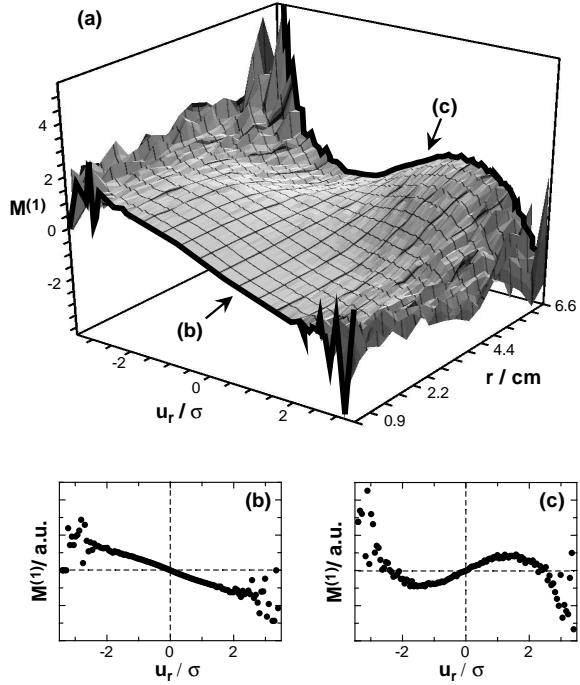, width=8.0cm}
  \end{center}
 \caption{Approximate Kramers Moyal coefficient $M^{(1)}(u_{r},r,\delta)$ for 
 $\delta =0.1$ mm at a distance of 8 d behind the cylinder (d=2 cm). b) and c) 
 corresponding $M^{(1)}$ for $r\approx 0.5$ cm and $r\approx 6.5$ cm. 
 Note the change of the sign of the slope of $M^{(1)}$ at $u_{r}=0$.}
 \label{KM3D}

\end{figure}
%
%
\begin{figure}[ht]
  \begin{center}
    \epsfig{file=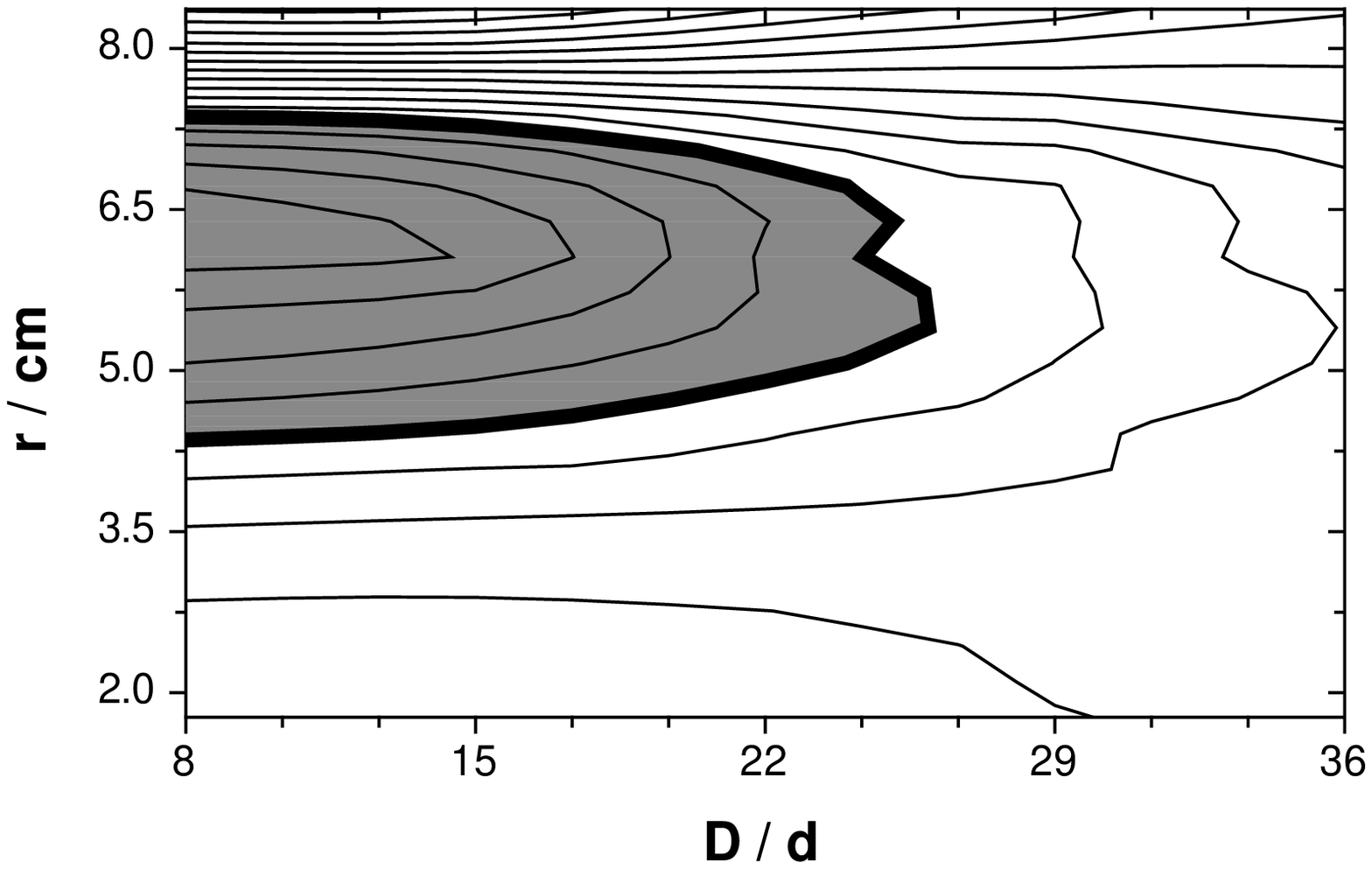, width=6.0cm}
  \end{center}
 \caption{Phase diagram for the transition to fully developed 
 turbulence in a flow behind a cylinder ($d=$2 cm), given by the 
 value of the slope $a_{1}$ of $M^{(1)}$ at $u_{r}=0$ (see figure
 \ref{KM3D}). Shadowy region corresponds to the occurrence of a 
 positive slope, i.e. the tendency to form the double 
 hump shape of the pfds (see figure \ref{pdfs}b).}
 \label{PhasDiagr}

\end{figure}
%

\begin{references}

\bibitem{turb} K.R. Sreenivasan, R.R. Antonia, Ann. Rev. Fluid 
Mech., {\bf29} 435 (1997); U. Frisch, Turbulence, Cambridge 1996


\bibitem{realturb} E. Gaudin, B. Protas, S. Goujon-Durand, J. Wojciechowski, 
J.E. Wesfreid, Phys. Rev. E, {\bf57}, R9 (1998); F. Chilla, J.F. 
Pinton, R. Labbe, Europhys. Lett. {\bf 35}, 271 (1996);
P.R. Van Slooten, Jayesh, S.B. Pope, Phys. Fluids {\bf 10},  246 
(1998); P. Olla, Phys. rev E {\bf 57}, 2824 (1998); .


\bibitem{Markov} R. Friedrich, J Peinke, Physica D {\bf 102} 147 (1997); 
                Phys.Rev.Lett. {\bf 78}, 863 (1997)

\bibitem{zeller} R. Friedrich, J. Zeller, J. Peinke, Europhys. Lett. 
                {\bf 41}, 153 (1998)

\bibitem{Risken} c.f.  P. H\"anggi and H. Thomas,
        Physics Reports {\bf 88}, 207 (1982); H. Risken, The Fokker-Planck equation,
        (Springer-Verlag Berlin, 1984).

\bibitem{scaling} J. Peinke, R. Friedrich, A. Naert, Z. Naturforsch 
                 {\bf 52 a}, 588 (1997).
\bibitem{windtunnel} Wind tunnel of the Lehrstuhl f\"ur 
Str\"omungsmechanik, University of Erlangen, 
Germany, was used.

\bibitem{Renner} Ch. Renner, B. Reisner, St. L\"uck, J. Peinke, R. Friedrich, 
chao-dyn/9811019


\bibitem{comment} It is known 
that the Chapman Kolmogorov equation is a necessary condition for the 
validity of a Markov process. There are only rare cases where the 
Chapman Kolmogorov equation holds when there is no Markov process 
present.

\bibitem{phasetr}  H. Haken, Synergetics (springen, Berlin 1983)

\bibitem{Ich}  St. L\"uck et. al. to be published.

\bibitem{Dyke} M. van Dyke, An Album of Fluid Motion (The Parabolic 
Press , Stanford 1982).
\end{references}
\end{document}